\documentclass[10pt,conference]{IEEEtran}
\IEEEoverridecommandlockouts
\usepackage[latin9]{inputenc}
\usepackage{geometry}
\usepackage{color}
\usepackage{float}
\usepackage{amsthm}
\usepackage{amsmath}
\usepackage{amssymb}
\usepackage{graphicx}

\providecommand{\tabularnewline}{\\}
\floatstyle{ruled}
\newfloat{algorithm}{tbp}{loa}
\providecommand{\algorithmname}{Algorithm}
\floatname{algorithm}{\protect\algorithmname}

\usepackage{color}
\usepackage{amsthm}
\usepackage{amsmath}
\usepackage{amssymb}
\usepackage{graphicx}
\usepackage{setspace}
\usepackage{url}

\usepackage{cite}
\usepackage{epsfig}
\usepackage{anysize}
\usepackage{algorithm}
\usepackage{algorithmic}
\usepackage{comment}
\usepackage{color}

\ifCLASSOPTIONcompsoc
\usepackage[caption=false,font=normalsize,labelfont=sf,textfont=sf]{subfig}
\else
\usepackage[caption=false,font=footnotesize]{subfig}
\fi

\theoremstyle{plain}
\newtheorem{thm}{\protect\theoremname}
\theoremstyle{plain}

\theoremstyle{plain}

\theoremstyle{plain}
\newtheorem{lem}{\protect\lemmaname}
\theoremstyle{remark}

\theoremstyle{note}

\marginsize{0.38in}{0.38in}{0.32in}{0.32in}

\makeatother

\providecommand{\notename}{Note}
\providecommand{\remarkname}{Remark}
\providecommand{\lemmaname}{Lemma}
\providecommand{\corollaryname}{Corollary}
\providecommand{\propositionname}{Proposition}
\providecommand{\theoremname}{Theorem}

\makeatother

\providecommand{\lemmaname}{Lemma}
\providecommand{\corollaryname}{Corollary}
\providecommand{\propositionname}{Proposition}
\providecommand{\remarkname}{Remark}
\providecommand{\theoremname}{Theorem}

\makeatother

\providecommand{\lemmaname}{Lemma}
\providecommand{\remarkname}{Remark}
\providecommand{\theoremname}{Theorem}

\IEEEoverridecommandlockouts
\begin{document}

\title{{Cache-Enabled Physical-Layer Security for Video Streaming in Wireless Networks with Limited Backhaul} \thanks{The work of D. W. K. Ng was supported by the Australian Research Council (ARC) Linkage Project LP 160100708. The work of R. Schober was supported by the Alexander von Humboldt Professorship Program. }\vspace{-.3cm} }
\author{{Lin Xiang, Derrick Wing Kwan Ng, Robert Schober, and Vincent W.S. Wong}\\
\vspace{-1cm}}
\author{\vspace{-.2cm}\IEEEauthorblockN{Lin Xiang$^*$, Derrick Wing Kwan Ng$^{\dag}$, Robert Schober$^*$, and Vincent W.S. Wong$^\ddag$}\\
\IEEEauthorblockA{$^*$Institute for Digital Communications, Friedrich-Alexander-Universit\"at Erlangen-N\"urnberg, Germany \\
$\dag$School of Electrical Engineering and Telecommunications, University of New South Wales, Australia \\
$\ddag$Department of Electrical and Computer Engineering, University of British Columbia, Vancouver, BC, Canada}
\vspace{-1cm}
}

\maketitle

\pagenumbering{gobble}
\begin{abstract}
In this paper, we investigate for the first time the benefits of wireless caching for the physical layer security (PLS) of wireless networks. In particular, a caching scheme enabling power-efficient PLS is proposed for cellular video streaming with constrained backhaul capacity. By sharing video data across a subset of base stations (BSs) through both caching and backhaul loading, secure cooperative transmission of several BSs is dynamically enabled in accordance with the cache status, the channel conditions, and the backhaul capacity. Thereby, caching reduces the data sharing overhead over the capacity-constrained backhaul links. More importantly, caching introduces additional secure degrees of freedom and enables a power-efficient design. We investigate the optimal caching and transmission policies for minimizing the total transmit power while providing quality of service (QoS) and guaranteeing secrecy during video delivery. A two-stage non-convex mixed-integer optimization problem is formulated, which optimizes the caching policy in an offline video caching stage and the cooperative transmission policy in an online video delivery stage. As the problem is NP-hard, suboptimal polynomial-time algorithms are proposed for low-complexity cache training and delivery control, respectively. Sufficient optimality conditions, under which the proposed schemes attain global optimal solutions, are also provided. Simulation results show that the proposed schemes achieve low secrecy outage probability and high power efficiency simultaneously. 
\end{abstract}

\section{Introduction}

The rapidly growing video-on-demand (VoD) streaming traffic in cellular networks has created significant challenges for cellular operators due to the scarce radio resources in the radio access network (RAN) and the limited capacity of the backhaul links \cite{WSJ13:Verizon}. 
To meet the stringent VoD streaming requirements in 5G cellular networks, wireless caching has been proposed in the literature 
\cite{Wang14:Cache,Liu14TSP:CoMP,PIMRC14:Bchlimit,Tao15Axiv:Multicast,Globecom15:CLCaching}.
In particular, caching has been exploited as a physical layer mechanism to facilitate traffic offloading on the backhaul, capacity enhancement and latency reduction in the RAN, and energy savings in the network.

Meanwhile, due to the broadcast nature of wireless transmission, VoD streaming data is vulnerable to potential eavesdroppers such as non-paying subscribers and malicious attackers. Secure video streaming schemes providing both video data protection and streaming quality of service (QoS) guarantee are thus preferred. Secure data delivery is not considered in cached-enabled transmission until recently \cite{Clancy15IT:Limits,AwanICC15:D2D,LandICC16:HetNets}. The existing works \cite{Clancy15IT:Limits,AwanICC15:D2D,LandICC16:HetNets} are motivated by the coded caching scheme proposed in \cite{Niesen14IT:CodedCaching}. Specifically, each user has a local cache to prestore parts of popular video content. By properly encoding (e.g. via network coding) the cached and the delivered contents, coded multicast delivery opportunities are enabled to achieve high delivery rates in serving various user requests \cite{Niesen14IT:CodedCaching}. In \cite{Clancy15IT:Limits}, a coded caching scheme is proposed to guarantee information delivery secrecy when eavesdroppers passively decipher the video data over the multicast link. For secrecy purpose, the cached and the delivered contents are encoded using random secret keys and \emph{secure} coded multicast delivery is enabled based on Shannon's one-time pad method. However, 
secure sharing of the secret keys can incur significant system overheads because the size of the secret keys should be large enough to keep the video file secret from the eavesdroppers. The coded caching scheme is extended to device-to-device (D2D) networks in \cite{AwanICC15:D2D}, where a sophisticated key generation and encryption scheme is investigated. Moreover, a secure delivery scheme, which prevents the eavesdroppers from obtaining the number of coded packets required for successful video file recovery, was proposed for cache-enabled heterogeneous small cell networks in \cite{LandICC16:HetNets}.

On the other hand, secure transmission has been thoroughly investigated for cellular networks. In particular, physical layer security (PLS) exploiting multi-input multi-output (MIMO) techniques has significant advantages over one-time pad based methods \cite{Khisti10IT:MISOME,Khisti10IT:MIMOME,LiTSP11:Opt-Robust}. For example, PLS techniques can opportunistically exploit the inherent randomness of wireless channels to enhance communication secrecy without using secret keys. In an $N_{\mathrm{t}}\times N_{\mathrm{r}}$ MIMO wiretap channel with full channel state information (CSI), information-theoretic studies have revealed that the secure degrees of freedom (s.d.o.f.)\footnote{Strictly positive s.d.o.f indicate  that the system's secrecy capacity can be scaled up by increasing the transmit power.} enabled by multiple antennas are given by $\min([N_{\mathrm{t}} - N_{\mathrm{e}}]^{+},N_{\mathrm{r}})$ \cite{Khisti10IT:MISOME,Khisti10IT:MIMOME}, where $N_{\mathrm{e}}$ is the number of eavesdropping antennas and $[x]^{+}=\max(x,0)$. Despite the increasing interest in secure cache-enabled communication, caching schemes facilitating PLS have not yet been reported in the literature.

To fill this void, this paper proposes a caching scheme for enhancing the PLS of cellular VoD streaming. Specifically, each base station (BS) is equipped with a cache. By caching the same video data across different BSs, more BSs can participate in the cooperative transmission of video data. Correspondingly, the s.d.o.f. can be significantly increased by exploiting the resulting large transmit antenna array \cite{Khisti10IT:MISOME}. Meanwhile, as caching reduces the data sharing overhead typically needed for cooperative transmission \cite{Liu14TSP:CoMP}, the s.d.o.f. are achievable even in cellular networks with capacity-constrained backhaul links. However, instead of trying to maximize the secrecy capacity, in this paper, we investigate the dual problem. Our goal is  \emph{to minimize the transmit power while satisfying delivery QoS and secrecy constraints}. Our work is inspired by \cite{Liu14TSP:CoMP,PIMRC14:Bchlimit,Tao15Axiv:Multicast}, which exploited cache-enabled cooperative transmission for transmit power minimization without secrecy considerations. Besides, backhaul capacity constraints are not considered in \cite{Liu14TSP:CoMP,PIMRC14:Bchlimit,Tao15Axiv:Multicast}. The main contributions of this paper are:
\begin{itemize}
\item We propose a caching scheme, which facilitates secure cellular video streaming with limited backhaul capacity. Thereby, the cache can reduce the backhaul traffic and support more BSs for secure cooperative transmission. 

\item We formulate a two-stage non-convex optimization problem to minimize the total BS transmit power subject to QoS and secrecy constraints. Efficient caching and delivery algorithms with polynomial time computational complexity are proposed to solve the problem, which are further shown to be globally optimal in certain regimes. 

\item Simulation results show that the proposed schemes can significantly enhance the PLS and reduce the total BS transmit power by efficiently utilizing the cache capacity.
\end{itemize}

Throughout this paper, $\mathbb{R}$ and $\mathbb{C}$ denote the sets of real and complex numbers, respectively; $\mathbf{I}_{L}$, $\mathbf{1}_{L}$, and $\mathbf{0}_{L}$ are the $L\times L$ identity, all-one, and zero matrices, respectively; $\mathsf{diag}(\mathbf{v})$ is a diagonal matrix with the diagonal elements given by $\mathbf{v}$; $(\cdot)^{T}$ and $(\cdot)^{H}$ are the transpose and complex conjugate transpose operators, respectively; $\mathsf{tr}(\cdot)$, $\mathsf{rank}(\cdot)$, $\det(\cdot)$, and $\lambda_{\max}(\cdot)$ denote the trace, rank, determinant, and maximum eigenvalue of a matrix, respectively; $\mathsf{Pr}(\cdot)$ denotes the probability mass operator; $\sim$ means distributed as; $|\mathcal{X}|$ represents the cardinality of set $\mathcal{X}$; $\mathcal{X}\times\mathcal{Y}$ denotes the Cartesian product of sets $\mathcal{X}$ and $\mathcal{Y}$; $\mathbf{A}\succeq\mathbf{0}$ ($\mathbf{A}\succ\mathbf{0}$) indicates that matrix $\mathbf{A}$ is positive semidefinite (definite); $\nabla_{\mathbf{X}}f\left(\mathbf{X}\right)$ denotes the complex-valued gradient of $f(\mathbf{X})$ with respect to $\mathbf{X}$; finally, $\left\lfloor \cdot\right\rfloor $ denotes the rounding operator. 

\section{\label{sec:System-Model}System Model}
%

\begin{figure}[t]
\vspace{-.4cm}
\centering
\includegraphics[scale=0.56]{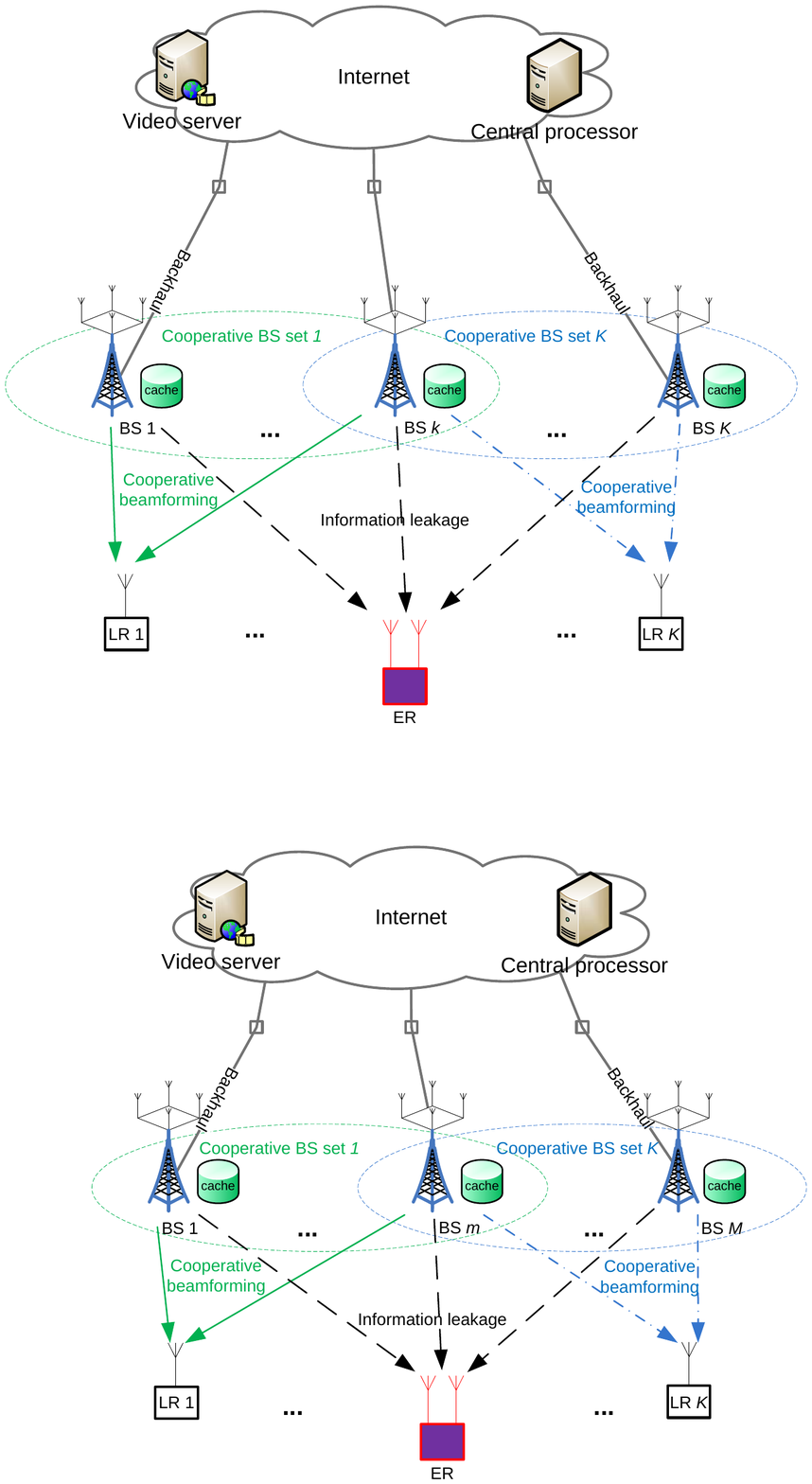}  
\protect\caption{\label{fig:System-Model}System model of cooperative beamforming for secure video delivery.}
\end{figure}

We consider video streaming in the downlink of a multi-cell cellular network as shown in Figure~\ref{fig:System-Model}. A set of BSs, $\mathcal{M}=\{1,\ldots,M\}$, each equipped with $N_{\mathrm{t}}$ antennas, broadcast the video data to a set of single-antenna legitimate receivers (LRs), $\mathcal{K}=\{1,\ldots,K\}$. A cache is deployed at each BS for prestoring the video data. Since the broadcast video data may be overheard by a passive eavesdropping receiver (ER), secure video delivery is applied to prevent potential information leakage to the ER. We assume that the ER is equipped with $N_{\mathrm{e}}$ antennas. 

The video server located on the Internet edge owns a library of video files, $\mathcal{F}=\{1,\ldots,F\}$, which are intended for delivery. The size of video file $f$ is $V_{f}$ bits. The BSs are connected to the video server via dedicated ``last-mile" backhaul links such as digital subscriber lines. We assume that the backhaul links are secure. However, since each backhaul is shared by different types of traffic (e.g. voice, data, multimedia, control signaling, etc.), the  backhaul capacity available for supporting  video streaming may be time-varying and limited \cite{Ng15TWC}. 

\subsection{Caching and BS Cooperation with Two-Stage Control}
The considered system is time-slotted. The system operation is divided into two stages. In the first stage, a portion of the video files is cached at the BSs, e.g. during the early mornings when cellular traffic is low. In the second stage, users request video files and a subset of the BSs cooperate to serve the requests. The video caching and delivery control decisions are determined at a central processor on the ``Internet edge'' and conveyed to the BSs via backhaul links.

We assume that file $f$ is split into  $L$ subfiles of equal size $V_f / L$ and each subfile $(f,l)\in\mathcal{F}\times \mathcal{L}$ is  delivered in one time slot, where $L \gg 1$ and $\mathcal{L} = \{1,\ldots,L\}$. Let binary variable $q_{f,l,m}\in\{0,1\}$ indicate the participation of BS $m\in\mathcal{M}$ in the cooperative transmission of subfile $(f,l)$. The set of  BSs cooperating in delivering subfile $(f,l)$ is then defined by $\mathcal{M}_{f,l}^{\mathrm{Coop}}\triangleq\left\{ m\in\mathcal{M}\mid q_{f,l,m}=1\right\} \subseteq\mathcal{M}$. To facilitate cooperative BS transmission, the video data can be conveyed to the  cooperating BSs in two manners: caching the data ahead of time or loading it via the backhaul links instantaneously during delivery. We propose a two-stage protocol for data caching and backhaul loading. Specifically, the data caching decisions are determined in the first stage based on the statistics or historical records of user requests, which are static. Since the information of the user requests, channel state, and backhaul capacity are only known online at the time of request, joint optimization of backhaul loading and BS cooperative transmission is deferred to the second stage when this information is available. 

Assuming that each subfile $(f,l)$ is encoded via rateless maximum distance separable (MDS) codes \cite{IT06Raptor},  we can cache a fraction of $c_{f,m}\in[0,1]$ and load via the backhaul a fraction of $b_{f,l,m}\in[0,1]$) of subfile $(f,l)$ at BS $m$. Here, $c_{f,m}$ does not have a subscript $l$ as we assume that the same portion of each subfile of file $f$ is cached at BS $m$. The relation between caching, backhaul loading, and cooperation formation is given by 
\begin{equation}
\textrm{C1: }\left\{ \begin{array}{c}
b_{f,l,m}=(1-c_{f,m})  q_{f,l,m}, \; f \in \mathcal{F},  m \in \mathcal{M,}\\
c_{f,m},b_{f,l,m}\in[0,1],\quad q_{f,l,m}\in\left\{ 0,1\right\}, \qquad
\end{array}\right.
\end{equation}
where cooperative transmission of subfile $(f,l)$ is possible, i.e., $q_{f,l,m}=1$, only when the subfile is fully available at BS $m$, i.e., $b_{f,l,m}+c_{f,m}=1$. Otherwise, C1 enforces $q_{f,l,m}=0$.

Let $C_{m}^{\max}$ and $B_{m}^{\max}$ be the cache capacity and the backhaul capacity available for video sharing at BS $m$, respectively. The cache placement and the backhaul loading are constrained by 
\begin{align}
\textrm{C2: } & \sum\nolimits_{f\in\mathcal{F}}c_{f,m}V_{f}\le C_{m}^{\max},\;  m\in\mathcal{M},\;\;\textrm{and}\\
\textrm{C3: } & \sum\nolimits_{f\in\mathcal{F}}b_{f,l,m}Q_{f}\le B_{m}^{\max},\; l \in \mathcal{L}, m\in\mathcal{M},\label{eq:C3}
\end{align}
respectively, where the fixed parameter $Q_{f}$ (in bps) represents the data rate required to load subfile $(f,l)$ at BS $m$. We have $Q_{f} = V_f / (\tau L )$ or equivalently $c_{f,m} {V_{f}}/{L} +  b_{f,l,m}Q_{f} \tau = {V_{f}}/{L}$, where $\tau$ denotes the duration of a  time slot. Moreover, $C_{m}^{\max}$ and $B_{m}^{\max}$ implicitly determine the number of cooperating BSs during online VoD streaming.

\subsection{Cooperative Beamforming for Secure Video Delivery}
When subfile $(f,l)$ is available at the subset of BSs  $ \mathcal{M}_{f,l}^{\mathrm{Coop}}$,
cooperative beamforming among these BSs is employed to deliver subfile $(f,l)$. Assume that an LR requests one (sub)file at a time. We denote a request from LR $k$ for subfile $(f,l)$ by $\boldsymbol{\rho}\triangleq(k, f, l)$ and the set of user requests by $\mathcal{S}\subseteq\mathcal{K}\times\mathcal{F}\times \mathcal{L}$. Here, $\mathcal{S}$ is known at the beginning of the online delivery stage. The source symbols of subfile $(f,l)$ in serving request $\boldsymbol{\rho}$ are denoted by $s_{\boldsymbol{\rho}}\in\mathbb{C}$, which are complex Gaussian random variables with $s_{\boldsymbol{\rho}}\sim\mathcal{CN}(0,\,1)$.

Let $\mathbf{w}_{m,\boldsymbol{\rho}}\in\mathbb{C}^{N_{\mathrm{t}}\times1}$ be the beamforming vector used at BS $m\in\mathcal{M}$ for sending symbol $s_{\boldsymbol{\rho}}$, where $\mathbf{w}_{m,\boldsymbol{\rho}}=\mathbf{0}$
if $m\notin\mathcal{M}_{f}^{\mathrm{Coop}}$. The joint transmit signal of BS set $\mathcal{M}$, denoted as $\mathbf{x}\in\mathbb{C}^{MN_{\mathrm{t}}\times1}$, is thus given by $\mathbf{x}  = \sum\nolimits_{\boldsymbol{\rho}\in\mathcal{S}}\mathbf{w}_{\boldsymbol{\rho}} {s}_{\boldsymbol{\rho}}$, where $\mathbf{w}_{\boldsymbol{\rho}}\triangleq[\mathbf{w}_{1,\boldsymbol{\rho}}^{H},\ldots,\mathbf{w}_{M,\boldsymbol{\rho}}^{H}]^{H}\in\mathbb{C}^{MN_{\mathrm{t}}\times1}$ is the joint beamforming vector for serving request $\boldsymbol{\rho}$. The beamforming vectors satisfy 
\begin{alignat}{1}
\textrm{C4: } & \mathsf{tr}\left(\boldsymbol{\Lambda}_{m}\mathbf{w}_{\boldsymbol{\rho}}\mathbf{w}_{\boldsymbol{\rho}}^{H}\right)\le q_{f,l,m}P_{m}^{\max},\; m\in\mathcal{M},\;\boldsymbol{\rho}\in\mathcal{S},\label{eq:coop3}\\
\textrm{C5: } & \mathsf{tr}\left(\sum\nolimits_{\boldsymbol{\rho}\in\mathcal{S}}\boldsymbol{\Lambda}_{m}\mathbf{w}_{\boldsymbol{\rho}}\mathbf{w}_{\boldsymbol{\rho}}^{H}\right)\le P_{m}^{\max},\; m\in\mathcal{M},\label{eq:perBS-txpwr}
\end{alignat}
where $P_{m}^{\max}$ is the maximum transmit power at BS $m$ and $\boldsymbol{\Lambda}_{m}$ is an  $MN_{\mathrm{t}}\times MN_{\mathrm{t}}$ diagonal matrix given by 
\begin{equation}
\boldsymbol{\Lambda}_{m}=\mathsf{diag}(\mathbf{0}_{(m-1)N_{\textrm{t}}\times1}^{T},\mathbf{1}_{N_{\textrm{t}}\times1}^{T},\mathbf{0}_{(M-m)N_{\textrm{t}}\times1}^{T}),
\end{equation}
i.e., $\mathsf{tr}\left(\mathbf{w}_{m,\boldsymbol{\rho}}\mathbf{w}_{m,\boldsymbol{\rho}}^{H}\right)=\mathsf{tr}\left(\boldsymbol{\Lambda}_{m}\mathbf{w}_{\boldsymbol{\rho}}\mathbf{w}_{\boldsymbol{\rho}}^{H}\right)$
holds. Herein, C5 limits the maximum transmit power per BS. C4 is a big-M constraint \cite{Floudas1995MINLP} on BS cooperation formation: if $q_{f,l,m}=0$ or  $b_{f,l,m} + c_{f,m} < 1$ (cf. C1), we have $\mathsf{tr}\left(\boldsymbol{\Lambda}_{m}\mathbf{w}_{\boldsymbol{\rho}}\mathbf{w}_{\boldsymbol{\rho}}^{H}\right)=\left\Vert \mathbf{w}_{m,\boldsymbol{\rho}}\right\Vert _{2}^{2}=0$, which results in $\mathbf{w}_{m,\boldsymbol{\rho}}=\mathbf{0}$; on the other hand, if $q_{f,l,m}=1$ and  $b_{f,l,m} + c_{f,m} = 1$, C4 is inactive due to C5. Thus, C4 enforces $\mathbf{w}_{m,\boldsymbol{\rho}}=\mathbf{0}$ whenever BS $m\notin\mathcal{M}_{f,l}^{\mathrm{Coop}}$ cannot participate in the cooperative transmission of subfile $(f,l)$.

We consider frequency flat fading channel during video transmission.  The received signals at LR\footnote{There is a one-to-one correspondence between requests and LRs. For convenience, the LRs are also indexed by $\boldsymbol{\rho}$ in the following when the requested (sub)files also need to be specified.} 
$\boldsymbol{\rho} \in \mathcal{S}$ and the ER, denoted by $y_{\boldsymbol{\rho}}\in\mathbb{C}$ and $\mathbf{y}_{\mathrm{e}} \in \mathbb{C}^{N_{\mathrm{e}}\times1}$, respectively, are given by 
\begin{alignat}{1}
y_{\boldsymbol{\rho}} & =\mathbf{h}_{\boldsymbol{\rho}}^{H}\mathbf{x}+z_{\boldsymbol{\rho}}\\
 & =\mathbf{h}_{\boldsymbol{\rho}}^{H}\mathbf{w}_{\boldsymbol{\rho}}s_{\boldsymbol{\rho}}+\sum\nolimits_{
 \boldsymbol{\rho}'\neq\boldsymbol{\rho}}\mathbf{h}_{\boldsymbol{\rho}}^{H}\mathbf{w}_{\boldsymbol{\rho}'}s_{\boldsymbol{\rho}'}+z_{\boldsymbol{\rho}},\; \boldsymbol{\rho}\in\mathcal{S},\label{eq:MUinterference}
\end{alignat}
and $\mathbf{y}_{\mathrm{e}}=\mathbf{G}^{H}\mathbf{x}+\mathbf{z}_{\mathrm{e}}$, where $\mathbf{h}_{\boldsymbol{\rho}} = [\mathbf{h}_{1,\boldsymbol{\rho}}^{H},\ldots,\mathbf{h}_{M,\boldsymbol{\rho}}^{H}]^{H}\in\mathbb{C}^{MN_{\mathrm{t}}\times1}$ and $\mathbf{G} =[\mathbf{G}_{1}^{H},\ldots,\mathbf{G}_{M}^{H}]^{H} \in \mathbb{C}^{MN_{\mathrm{t}} \times N_{\mathrm{e}}}$ are the channel matrix from BS set $\mathcal{M}$ to LR $\boldsymbol{\rho}$ and the ER, respectively. $\mathbf{h}_{m,\boldsymbol{\rho}} \in \mathbb{C}^{N_{\mathrm{t}} \times 1}$ and $\mathbf{G}_{m} \in \mathbb{C}^{N_{\mathrm{t}} \times N_{\mathrm{e}}}$ model the channels between BS $m\in\mathcal{M}$ and the corresponding LR/ER receivers; and $z\mathbf{_{\boldsymbol{\rho}}} \sim \mathcal{CN}(0,\sigma^{2})$ and $\mathbf{z}_{\mathrm{e}} \sim \mathcal{CN}(\mathbf{0},\sigma_{\mathrm{e}}^{2} \mathbf{I}_{N_{\mathrm{e}}})$ are the zero-mean complex Gaussian noises at the LRs and the ER  with variance $\sigma^2$ and covariance matrix $\sigma_{\mathrm{e}}^{2}\mathbf{I}_{N_{\mathrm{e}}}$, respectively. The achievable rate at LR $\boldsymbol{\rho}$, denoted by $R_{\boldsymbol{\rho}}$, is given by 
\begin{align}
R_{\boldsymbol{\rho}} & =\log\left(1+\Gamma_{\boldsymbol{\rho}}\right),\quad\boldsymbol{\rho}\in\mathcal{S},\label{eq:LR-rate}\\
\Gamma_{\boldsymbol{\rho}} & =\frac{\frac{1}{\sigma^{2}}\left|\mathbf{h}_{\boldsymbol{\rho}}^{H}\mathbf{w}_{\boldsymbol{\rho}}\right|^{2}}{1+\frac{1}{\sigma^{2}}\sum_{\boldsymbol{\rho}'\in\mathcal{S},\boldsymbol{\rho}'\neq\boldsymbol{\rho}}\left|\mathbf{h}_{\boldsymbol{\rho}}^{H}\mathbf{w}_{\boldsymbol{\rho}'}\right|^{2}},
\end{align}
where $\Gamma_{\boldsymbol{\rho}}$ is the received signal-to-interference-plus-noise ratio (SINR) at LR $\boldsymbol{\rho}$.

We assume that under a worst-case scenario the ER can eavesdrop the information intended for each LR after canceling the interference caused by all other LRs. This is possible if the ER adopts advanced receiver structures such as successive interference cancellation decoders \cite{Tse2005Fundamentals}. For guaranteeing secure VoD streaming, the proposed secure delivery scheme is designed to avoid VoD data leakage even in such a worst-case scenario. Thus, an achievable secrecy rate for LR $\boldsymbol{\rho}$ is given by \cite{Khisti10IT:MISOME,LiTSP11:Opt-Robust}
\begin{equation}
\begin{aligned}R_{\boldsymbol{\rho}}^{\mathrm{sec}} & =\left[R_{\boldsymbol{\rho}}-R_{\mathrm{e},\boldsymbol{\rho}}\right]^{+},\quad\boldsymbol{\rho}\in\mathcal{S},\end{aligned}
\end{equation}
where $R_{\mathrm{e},\boldsymbol{\rho}}$ denotes the capacity of the ER in decoding LR $\boldsymbol{\rho}$ for subfile $(f,l)$ and is given by
\begin{equation}
\begin{aligned}R_{\mathrm{e},\boldsymbol{\rho}} & =\log\det\left(\mathbf{I}_{N_{\textrm{e}}}+\frac{1}{\sigma_{\textrm{e}}^{2}}\mathbf{G}\mathbf{G}^{H}\mathbf{w}_{\boldsymbol{\rho}}\mathbf{w}_{\boldsymbol{\rho}}^{H}\right),\;\boldsymbol{\rho}\in\mathcal{S}.\end{aligned}
\end{equation}

\section{\label{sec:Problem-Formulation}Two-Stage Problem Formulation}
In this paper, we assume that the ER is a non-paying video subscriber and the CSIs of all the subscribers are perfectly known\footnote{Under the perfect CSI assumption, artificial noise based jamming methods are suboptimal for the considered system \cite{LiTSP11:Opt-Robust} and thus not considered in this paper.}. A two-stage optimization problem is then formulated to minimize the total BS transmit power. Specifically, in the first stage, the cached video data is optimized offline. In the second stage, the cooperative transmission strategies are optimized online for given cache and backhaul status. We note that the results in this paper provide a performance upper bound for the case of imperfect CSI.

\vspace{-.2cm}
\subsection{Second-Stage Online Delivery Control }
The BS cooperation formation policy $\{q_{f,l,m},\, b_{f,l,m}\}$ and the cooperative transmission policy $\{\mathbf{w}_{\boldsymbol{\rho}}\}$ are optimized in the second stage. For this purpose, we assume that the set of user requests $\mathcal{S}$ is given and the cache status $\left\{ c_{f,m}\right\} $ has already been determined in the first stage. Let $\mathbf{D}_{\text{\mbox{II}}} \triangleq [q_{f,l,m},\, b_{f,l,m},\mathbf{w}_{\boldsymbol{\rho}}]$ be the second-stage (delivery) optimization space. The second-stage problem is formulated as follows, 
\begin{align}
\textrm{R0:}\quad\min_{\mathbf{D}_{\text{\mbox{II}}}}\quad & f_{\mathrm{\text{\mbox{II}}}}\triangleq\sum\nolimits_{\boldsymbol{\rho}\in\mathcal{S}}\mathsf{tr}(\mathbf{w}_{\boldsymbol{\rho}}\mathbf{w}_{\boldsymbol{\rho}}^{H})\label{eq:R1}\\
\mathrm{\textrm{s.t. \quad}} & \textrm{C1, C3, C4},\textrm{ C5},\nonumber \\
 & \textrm{QoS constraint C6: }R_{\boldsymbol{\rho}}\ge R_{\boldsymbol{\rho}}^{\textrm{req}},\;\boldsymbol{\rho}\in\mathcal{S},\nonumber \\
 & \textrm{Secrecy constraint C7: }R_{\mathrm{e},\boldsymbol{\rho}}\le R_{\mathrm{e},\boldsymbol{\rho}}^{\mathrm{tol}},\;\boldsymbol{\rho}\in\mathcal{S},\nonumber 
\end{align}
where C6 guarantees a minimum video delivery rate, $R_{\boldsymbol{\rho}}^{\textrm{req}}$, to provide streaming QoS for LR $\boldsymbol{\rho}$.  C7 restricts the capacity of the ER to be below a maximum tolerable secrecy threshold $R_{\mathrm{e},\boldsymbol{\rho}}^{\mathrm{tol}}$ for video data protection. Note that C6 and C7 together guarantee a minimum achievable secrecy rate of $R_{\boldsymbol{\rho}}^{\mathrm{sec}}=[R_{\boldsymbol{\rho}}^{\textrm{req}}-R_{\mathrm{e},\boldsymbol{\rho}}^{\mathrm{tol}}]^{+}$ for LR $\boldsymbol{\rho}$.

\subsection{First-Stage Offline Cache Training}
A historical data driven approach~\cite{Globecom15:CLCaching} 
 is adopted for the \emph{offline} caching scheme in the first stage. Assume that $\Omega$ sets of scenario data are available for training the cache, each set consisting of user requests, CSI, and the available backhaul capacities at a particular time instant. The scenario data is indexed by $\omega\in\left\{ 1,\ldots,\Omega\right\} $. Let $\mathbf{C}_{\text{\mbox{I}}}\triangleq[c_{f,m},\,\mathbf{D}_{\text{\mbox{I}},\omega}]$ be the first-stage (caching) optimization space, where $\mathbf{D}_{\text{\mbox{I}},\omega}\triangleq[q_{f,l,m,\omega},b_{f,l,m,\omega},\mathbf{w}_{\boldsymbol{\rho},\omega}]$ denotes the auxiliary delivery decisions for scenario $\omega$ during training. We define the feasible delivery set for scenario $\omega$ by $\mathbf{\mathcal{D}}_{\text{\mbox{I}},\omega}\triangleq\left\{ \mathbf{D}_{\text{\mbox{I}},\omega}\mid\textrm{C1, C4--C7}\right\} $, where C1 and C4--C7 need to be reformulated with an augmented system state space. For example, C1 is rewritten as 
\begin{equation}
\textrm{C1: }\left\{ \begin{array}{c}
b_{f,l,m,\omega}=(1-c_{f,m})  q_{f,l,m,\omega}, \qquad \\
c_{f,m},b_{f,l,m,\omega}\in[0,1],\;\; q_{f,l,m,\omega}\in\left\{ 0,1\right\} ,
\end{array}\right.
\label{C1-2}
\end{equation}
and C4--C7 are similarly formulated.

The first stage problem is then formulated to minimize the average transmit power for the considered scenarios, i.e.,
\begin{align}
\textrm{Q0:}\quad\min_{\mathbf{C}_{\text{\mbox{I}}}}\quad & \frac{1}{\Omega}\sum\nolimits_{\omega=1}^{\Omega}\; f_{\mathrm{\text{\mbox{I}}},\omega} 
\label{eq-Q0}\\
\mathrm{s.t.}\quad & \textrm{C2, }\ensuremath{\overline{\textrm{C3}}},\;\mathbf{D}_{\text{\mbox{I}},\omega}\in\mathbf{\mathcal{D}}_{\text{\mbox{I}},\omega},\;\omega\in\left\{ 1,\ldots,\Omega\right\} ,\nonumber 
\end{align}
where $f_{\text{\mbox{I}},\omega}\triangleq\sum_{\boldsymbol{\rho}\in\mathcal{S}}\mathsf{tr}(\mathbf{w}_{\boldsymbol{\rho},\omega}\mathbf{w}_{\boldsymbol{\rho},\omega}^{H})$
is the instantaneous transmit power for scenario $\omega$. {{The objective function in \eqref{eq-Q0} is the empirical average of the transmit powers of all scenarios.}} $\ensuremath{\overline{\textrm{C3}}}$
is an average backhaul capacity constraint given by 
\[
\ensuremath{\overline{\textrm{C3}}}:\frac{1}{\Omega}\sum\nolimits_{\omega=1}^{\Omega}\sum\nolimits_{f\in\mathcal{F}}b_{f,l,m,\omega}Q_{f}\le\frac{1}{\Omega}\sum\nolimits_{\omega=1}^{\Omega}B_{m,\omega}^{\max}, \, m\in\mathcal{M},
\]
which is a relaxation of the per-scenario backhaul capacity constraints $\sum_{f\in\mathcal{F}}b_{f,l,m,\omega}Q_{f}\le B_{m,\omega}^{\max}$, $\omega\in\left\{ 1,\ldots,\Omega\right\} $. In the considered two-stage control, $\ensuremath{\overline{\textrm{C3}}}$ avoids the conservative use of the backhaul links in the first stage when the actual backhaul capacity at the time of delivery is uncertain; instead, the actual cooperative transmission decisions are flexibly deferred to the second stage when the available backhaul capacity is known online. Furthermore, $\ensuremath{\overline{\textrm{C3}}}$ leads to computational advantages for low complexity cache training as revealed in Section \ref{sec4-3}.

Both R0 and Q0 are non-convex mixed-integer nonlinear programs (MINLPs)%
\footnote{For a non-convex MINLP, even if the integer constraints are relaxed to convex constraints, the problem remains non-convex \cite{Floudas1995MINLP}.} 
due to non-convex constraints C6, C7, and binary optimization variables $q_{f,l,m}\in\left\{ 0,1\right\} $ and $q_{f,l,m,\omega} \in \left\{ 0,1\right\}$. Moreover, Q0 involves bilinear constraint C1. This type of problem is generally NP-hard and there are no known polynomial time algorithms to  solve them optimally \cite{Floudas1995MINLP}. To strike a balance between computational complexity and optimality, we present two effective polynomial time suboptimal algorithms for solving R0 and Q0 in Section \ref{sec:Problem-Solution}. The proposed algorithms become optimal when the cache capacity and the number of scenarios are sufficiently large, respectively.

\section{\label{sec:Problem-Solution}Problem Solution}
In this section, the solutions of Problems R0 and Q0 are presented. We start by solving Problem R0. The solution method employed for solving R0 is then extended to tackle Q0. 

\vspace{-.2cm}
\subsection{\label{sub4-1}Optimal Solution of R0 in Large Cache Capacity Regime}
We first discuss special conditions under which Problem R0 is polynomial time solvable. The results derived herein also shed light on how to solve R0 for the general case in Section \ref{sub4-2}.

Let $\mathcal{F}(\mathcal{S})\triangleq\left\{ f\mid(\cdot,f,\cdot)\in\mathcal{S}\right\} $ be the set of files requested by $\mathcal{S}$, where $\mathcal{F}(\mathcal{S})\subseteq\mathcal{F}$. We define $F(\mathcal{S})\triangleq\left|\mathcal{F}(\mathcal{S})\right|$, which satisfies $F(\mathcal{S})\le\min\left\{ \left|\mathcal{S}\right|,F\right\} $. For given $\left\{ c_{f,m}\right\} $, the backhaul loading decision variables $b_{f,l,m}$ can be eliminated based on C1. As a result, Problem R0
is reformulated as
\begin{align}
\textrm{R0:}\quad\min_{\mathbf{D}_{\text{\mbox{II}}}}\quad & f_{\mathrm{\text{\mbox{II}}}}\label{eq:R1-1}\\
\mathrm{\textrm{s.t. \quad}} & \textrm{C4,}\textrm{ C5, C6, C7, } 
 \overline{\textrm{C1}}:\, q_{f,l,m}\in\left\{ 0,1\right\} ,\nonumber \\
 & \widetilde{\textrm{C3}}:\sum\nolimits_{f\in\mathcal{F}(\mathcal{S})}q_{f,l,m}Q_{f,m}\le B_{m}^{\max},\; m\in\mathcal{M},\nonumber 
\end{align}
where $Q_{f,m}\triangleq Q_{f}\times(1-c_{f,m})$ is the ``effective'' data rate on the backhaul link for loading subfile $(f,l)$ into BS $m$ and hence, $\widetilde{\textrm{C3}}$ and $\textrm{C3}$ are equivalent. We have the following lemma regarding the BS cooperation formation in \eqref{eq:R1-1}. 
\vspace{-.2cm}
\begin{lem}
\label{lem:(Monotonicity-of-R0)}\emph{(Monotonicity of R0). For cooperation
sets $\mathcal{M}_{f,l}^{\mathrm{Coop,1}}\subseteq\mathcal{M}_{f,l}^{\mathrm{Coop,2}},\,\forall (f,l) \in \mathcal{F}(\mathcal{S}) \times \mathcal{L}$,
the corresponding optimal cooperative transmission powers, denoted
by $f_{\mathrm{\text{\mbox{II}}}}^{1}$, $f_{\mathrm{\text{\mbox{II}}}}^{2}$,
respectively, satisfy $f_{\mathrm{\text{\mbox{II}}}}^{1}\ge f_{\mathrm{\text{\mbox{II}}}}^{2}$. }
\end{lem}
\vspace{-.1cm}
\begin{IEEEproof} {{If $\mathcal{M}_{f,l}^{\mathrm{Coop},i}$ is adopted for solving R0,   let $\mathcal{D}_{\boldsymbol{\rho}}^{{i}}$ and $\mathcal{D}_{m,\boldsymbol{\rho}}^{{i}}$ be the resulting feasible sets of $\mathbf{w}_{\boldsymbol{\rho}}$ and $\mathbf{w}_{m,\boldsymbol{\rho}}$, respectively, where $\mathcal{D}_{\boldsymbol{\rho}}^{{i}} = \prod_{m=1}^M \mathcal{D}_{m,\boldsymbol{\rho}}^{{i}}$, ${i}=1,2$. Considering C4, we have $\mathbf{0} \in \mathcal{D}_{m,\boldsymbol{\rho}}^{{i}} $ if $m \in \mathcal{M}$, and $\mathcal{D}_{m,\boldsymbol{\rho}}^{{i}} = \{ \mathbf{0} \} $ if $ m \notin \mathcal{M}_{f,l}^{\mathrm{Coop},i}$. Besides, $\mathcal{D}_{m,\boldsymbol{\rho}}^{\mathrm{1}} = \mathcal{D}_{m,\boldsymbol{\rho}}^{\mathrm{2}}$ if $m \in \mathcal{M}_{f,l}^{\mathrm{Coop, 1}}$ and $m \in \mathcal{M}_{f,l}^{\mathrm{Coop, 2}}$. Thus, $\mathcal{D}_{\boldsymbol{\rho}}^{\mathrm{1}} \subseteq \mathcal{D}_{\boldsymbol{\rho}}^{\mathrm{2}}$ holds if 
$\mathcal{M}_{f,l}^{\mathrm{Coop,1}}\subseteq\mathcal{M}_{f,l}^{\mathrm{Coop,2}}$. Since the objective function of R0 is only a function of $\mathbf{w}_{\boldsymbol{\rho}}$, we have $f_{\mathrm{\text{\mbox{II}}}}^{1} \ge  f_{\mathrm{\text{\mbox{II}}}}^{2}$, which completes the proof. 
%
}}
\end{IEEEproof}
Based on Lemma \ref{lem:(Monotonicity-of-R0)}, the backhaul capacity constraints C3 or $\widetilde{\textrm{C3}}$ can be removed in systems with large cache capacity and there is no loss of optimality since the cache can effectively offload the backhaul traffic. That is, fully cooperative transmission with cooperative set $\mathcal{M}_{f,l}^{\mathrm{F-Coop}}=\mathcal{M}$ is optimal since $\mathcal{M}_{f,l}^{\mathrm{Coop}}\subseteq\mathcal{M}_{f,l}^{\mathrm{F-Coop}},\forall\mathcal{M}_{f,l}^{\mathrm{Coop}}$.
Furthermore, Problem R0 is polynomial time solvable. This result holds whenever the cooperation sets are fixed. Without loss of generality, we show now that the resulting problem R0($\mathbf{w}_{\boldsymbol{\rho}}$), i.e.,   
\begin{align}
\textrm{R0(\ensuremath{\mathbf{w}_{\boldsymbol{\rho}}}):}\quad\min_{\mathbf{w}_{\boldsymbol{\rho}}}\quad & f_{\mathrm{\text{\mbox{II}}}}\\
\mathrm{\textrm{s.t. \quad}} & \textrm{C4, C5, C6, C7,}\nonumber
\end{align}
where $\mathbf{w}_{\boldsymbol{\rho}}$ is the optimization variable and the cooperation formation decisions are known a priori, is polynomial time solvable. In particular, although R0($ \mathbf{w}_{\boldsymbol{\rho}} $) is non-convex due to the non-convex constraints C6 and C7, we reveal the hidden convexity of  R0($\mathbf{w}_{\boldsymbol{\rho}}$) by problem transformation.

Let $\mathbf{W}_{\boldsymbol{\rho}} \triangleq \mathbf{w}_{\boldsymbol{\rho}} \mathbf{w}_{\boldsymbol{\rho}}^{H} \succeq \mathbf{0}$ and $\mathbf{H}_{\boldsymbol{\rho}} \triangleq \mathbf{h}_{\boldsymbol{\rho}} \mathbf{h}_{\boldsymbol{\rho}}^{H}$. The QoS constraint C6 can be transformed into affine constraints, 
\begin{equation*}
\begin{aligned}\textrm{C6}\iff & \Gamma_{\boldsymbol{\rho}}\ge\kappa_{\boldsymbol{\rho}}^{\mathrm{req}}\triangleq2^{R_{\boldsymbol{\rho}}^{\mathrm{req}}}-1,\\
\quad\,\iff & \overline{\textrm{C6}}:\frac{1}{\kappa_{\boldsymbol{\rho}}^{\mathrm{req}}}\mathsf{tr}\left(\mathbf{W}_{\boldsymbol{\rho}}\mathbf{H}_{\boldsymbol{\rho}}\right)\ge\sigma^{2}
+\sum\nolimits_{\boldsymbol{\rho}'\neq\boldsymbol{\rho}}\mathsf{tr}\left(\mathbf{W}_{\boldsymbol{\rho}'}\mathbf{H}_{\boldsymbol{\rho}}\right).
\end{aligned}
\label{eq:qos-con-trans1}
\end{equation*}
C6 and $\overline{\textrm{C6}}$ are equivalent if and only if the following constraint holds 
\[\textrm{C8:} \; \mathbf{W}_{\boldsymbol{\rho}}\succeq\mathbf{0} \textrm{ and } \mathsf{rank}(\mathbf{W}_{\boldsymbol{\rho}})\le1.\]
Furthermore, the following lemma is needed to transform the security constraint C7.

\vspace{-.2cm}
\begin{lem}[\kern-.3em\cite{LiTSP11:Opt-Robust}]
\label{lem:det-tr-ineq}
\emph{For $\mathbf{A}\succeq\mathbf{0}$, we have 
\begin{equation}
\det(\mathbf{I}+\mathbf{A})\ge1+\mathsf{tr}(\mathbf{A}),\label{eq:det-tr-ineq1}
\end{equation}
where the equality holds if and only if $\mathsf{rank}(\mathbf{A})\le1$.}
\end{lem}

Based on Lemma \ref{lem:det-tr-ineq}, we have
\begin{alignat*}{1}
\textrm{C7}\stackrel{\textrm{(a)}}{\iff} & \;\; \det\left(\mathbf{I}_{N_{\textrm{e}}}+\frac{1}{\sigma_{\textrm{e}}^{2}}\mathbf{G}^{H}\mathbf{W}_{\boldsymbol{\rho}}\mathbf{G}\right)\le2^{R_{\mathrm{e},\boldsymbol{\rho}}^{\mathrm{tol}}}\\
\stackrel{\textrm{(b)}}{\Longrightarrow} & \quad\mathsf{tr}\left(\mathbf{G}^{H}\mathbf{W}_{\boldsymbol{\rho}}\mathbf{G}\right)\le\sigma_{\textrm{e}}^{2}\kappa_{\boldsymbol{\rho}}^{\mathrm{tol}}\triangleq\sigma_{\textrm{e}}^{2}\left(2^{R_{\mathrm{e},\boldsymbol{\rho}}^{\mathrm{tol}}}-1\right) \nonumber \\
\stackrel{\textrm{(c)}}{\Longrightarrow} & \quad\lambda_{\max}\left(\mathbf{G}^{H}\mathbf{W}_{\boldsymbol{\rho}}\mathbf{G}\right) \le\sigma_{\textrm{e}}^{2}\kappa_{\boldsymbol{\rho}}^{\mathrm{tol}} \nonumber \\
\Longleftrightarrow & \quad\textrm{\ensuremath{\overline{\textrm{C7}}}:}\quad\mathbf{G}^{H}\mathbf{W}_{\boldsymbol{\rho}}\mathbf{G}\preceq\sigma_{\textrm{e}}^{2}\kappa_{\boldsymbol{\rho}}^{\mathrm{tol}}\mathbf{I},\quad\boldsymbol{\rho}\in\mathcal{S},\nonumber 
\end{alignat*}
where (a) follows from $\det(\mathbf{I}+\mathbf{A}\mathbf{B})=\det(\mathbf{I}+\mathbf{B}\mathbf{A})$,
(b) follows from \eqref{eq:det-tr-ineq1}, and (c) is due to the inequality $\lambda_{\max}(\mathbf{A})\le\mathsf{tr}(\mathbf{A})$ for $\mathbf{A}\succeq\mathbf{0}$. For (b) and (c), equality holds if $\mathsf{rank}(\mathbf{A})\le1$. Therefore, C7 and $\overline{\textrm{C7}}$ are equivalent if  $\mathsf{rank}(\mathbf{W}_{\boldsymbol{\rho}})\le1$.

By applying the above transformations and relaxing the rank constraint $\mathsf{rank}(\mathbf{W}_{\boldsymbol{\rho}})\le1$, i.e., removing it from C8, we obtain the following convex semidefinite program (SDP), 
\begin{align}
\textrm{R1:} \quad\min_{\mathbf{W}_{\boldsymbol{\rho}}}\quad & \mathsf{tr} \left(\sum\nolimits_{\boldsymbol{\rho}\in\mathcal{S}}\mathbf{W}_{\boldsymbol{\rho}} \right)\\
\mathrm{\textrm{s.t. \quad}} & \textrm{\ensuremath{\overline{\textrm{C4}}}: }\mathsf{tr}\left(\boldsymbol{\Lambda}_{m}\mathbf{W}_{\boldsymbol{\rho}}\right)\le q_{f,l,m}P_{m}^{\max},\; m\in\mathcal{M},\nonumber \\ 
 & \textrm{\ensuremath{\overline{\textrm{C5}}}: }\mathsf{tr}\left(\sum\nolimits_{\boldsymbol{\rho}}\boldsymbol{\Lambda}_{m}\mathbf{W}_{\boldsymbol{\rho}}\right)\le P_{m}^{\max},\; m\in\mathcal{M},\nonumber \\
 & \textrm{\ensuremath{\overline{\textrm{C6}}}, \ensuremath{\overline{\textrm{C7}}}, } \textrm{\ensuremath{\overline{\textrm{C8}}}: } \mathbf{W}_{\boldsymbol{\rho}}\succeq\mathbf{0},\;\boldsymbol{\rho}\in\mathcal{S}.\nonumber 
\end{align}
Generally, Problem R1 achieves a lower bound on the optimal value of R0($\mathbf{w}_{ \boldsymbol{\rho}}$). However, if the solution of R1 also satisfies $\mathsf{rank}(\mathbf{W}_{\boldsymbol{\rho}}^{*}) \le 1,\;\boldsymbol{\rho}\in\mathcal{S}$,
then $\mathbf{W}_{\boldsymbol{\rho}}^{*}$ is also the optimal solution of Problem R0($\mathbf{w}_{\boldsymbol{\rho}}$), i.e., the relaxation is tight. For the problem at hand, the relaxation in R1 is always tight, which is established in the following theorem.

\vspace{-.2cm}
\begin{thm}
\label{prop2}\emph{Problems R0($\mathbf{w}_{\boldsymbol{\rho}}$) and R1 are equivalent} \emph{in the sense that both problems have the same optimal value; in particular, the optimal solution $\mathbf{W}_{\boldsymbol{\rho}}^{*}$ of R1 satisfies $\mathsf{rank}(\mathbf{W}_{\boldsymbol{\rho}}^{*}) \le 1,\;\boldsymbol{\rho}\in\mathcal{S}$,
and the optimal beamforming vector $\mathbf{w}_{\boldsymbol{\rho}}^{*}$ of R0($\mathbf{w}_{\boldsymbol{\rho}}$) is given by the principal eigenvector of $\mathbf{W}_{\boldsymbol{\rho}}^{*}$. }
\end{thm}
\begin{IEEEproof}
Please refer to Appendix \ref{append2}. 
\end{IEEEproof}
Based on Theorem~\ref{prop2}, Problem R0($\mathbf{w}_{\boldsymbol{\rho}}$) can be efficiently solved with convex optimization algorithms. For example, the interior-point method \cite{Bertsekas82Lagrange}, which has been implemented in numerical solvers such as CVX \cite{cvx}, is applicable.

\vspace{-.3cm}
\subsection{\label{sub4-2}General Case: Greedy Iterative Solution of R0}
In regimes with limited cache capacity, Problem R0 is NP-hard due to binary constraint C1. In particular, the optimal solution has to be determined by enumerating all possible cooperation sets satisfying backhaul capacity constraint C3. For this purpose, enumeration methods such as exhaustive search and branch-and-bound \cite{Floudas1995MINLP} are applicable. However, although the remaining cooperative beamforming problem, i.e., Problem R0($\mathbf{w}_{\boldsymbol{\rho}}$), can be efficiently solved for each choice of cooperation sets, cf. Theorem \ref{prop2}, the overall computational complexity grows exponentially with the number of BSs. To be specific, we define $\overline{T}_{m}\triangleq\min\left\{ \left\lfloor B_{m}^{\max}/\min_{f \in \mathcal{F}(\mathcal{S})} Q_{f,m} \right\rfloor ,F(\mathcal{S})\right\} $ and $\underline{T}_{m} \triangleq \left\lfloor B_{m}^{\max}/\max_{f\in\mathcal{F}(\mathcal{S})}Q_{f,m}\right\rfloor $.
According to Lemma \ref{lem:(Monotonicity-of-R0)}, the optimal cooperation formation solutions are contained in the vertices of polyhedral simplexes defined by $\sum_{f\in\mathcal{F}(\mathcal{S})}q_{f,l,m}\le T_{m}$ and $q_{f,l,m}\in[0,1]$, where $\underline{T}_{m}\le T_{m}\le\overline{T}_{m},m\in\mathcal{M}$. As a result, an enumeration over approximately $\prod_{m=1}^{M}\left(_{F(\mathcal{S})}^{T_{m}}\right)$ choices of the cooperation sets is required in total (in the worst case) for solving R0 by exhaustive search (branch-and-bound).

Non-polynomial time enumeration methods are only applicable for small systems. For practical large systems, however, effective polynomial time algorithms are preferred. Herein, a low-complexity iterative algorithm is proposed based on greedy heuristics to solve R0 and summarized in Algorithm~\ref{alg1}.

Let $k$ be the iteration index. Define $\mathcal{Q}_{k}\triangleq\left\{ (f,m)\mid q_{f,l,m}=1,m\in\mathcal{M},f\in F(\mathcal{S})\right\} $ as the BS cooperation solution set determined at iteration $k$. Algorithm \ref{alg1} starts with the initialization $\mathcal{Q}_{0}=\prod_{f\in\mathcal{F}(\mathcal{S})}\mathcal{M}_{f,l}^{\mathrm{F-Coop}}=\prod_{f\in\mathcal{F}(\mathcal{S})}\mathcal{M}$.
At iteration $k=1,2,\ldots$, the values of $q_{f,l,m}$ are fixed according to $\mathcal{Q}_{k-1}$ and the remaining cooperative beamforming solutions are solved via Problem R0($\mathbf{w}_{\boldsymbol{\rho}}$), whose optimal value is denoted by $f_{\mathrm{\text{\mbox{II}}}}^{*}(\mathcal{Q}_{k-1})$. If $\mathcal{Q}_{k-1}$ fulfills the backhaul capacity constraint C3, the algorithm stops and returns the solutions for cooperative BS transmission. Otherwise, the greedy algorithm sets $q_{f',l,m'}=0$ for $(f',m')\in\mathcal{Q}_{k-1}$ which has the least penalty on the total transmit power, i.e.,
\begin{alignat}{1}
(f',m') & \in\underset{(f,m)\in\mathcal{F}(\mathcal{S})\times\mathcal{M}_{k}^{\mathrm{vio}}}{\arg\min} \! \left[f_{\mathrm{\text{\mbox{II}}}}^{*}(\mathcal{Q}_{k-1}\backslash\left\{ (f,m)\right\} )-f_{\mathrm{\text{\mbox{II}}}}^{*}(\mathcal{Q}_{k-1})\right],\label{eq:greedy}\\
\quad\;\;\mathcal{Q}_{k} & =\mathcal{Q}_{k-1}\backslash\left\{ (f',m')\right\} ,
\end{alignat}
where $\mathcal{M}_{k-1}^{\mathrm{vio}}$ denotes the index set of BSs violating constraint C3 if $\mathcal{Q}_{k-1}$ is adopted for cooperative transmission, i.e., 
\begin{equation}
\mathcal{M}_{k-1}^{\mathrm{vio}}\triangleq  \Big\{ m\in\mathcal{M} \mid \sum\nolimits_{(f,m) \in \mathcal{Q}_{k-1}}Q_{f,m}>B_{m}^{\max}  \Big\}.\label{eq:vioBS}
\end{equation}
The iteration process is repeated until C3 is fulfilled.

Note that during each iteration of Algorithm \ref{alg1}, \eqref{eq:greedy} is solved by enumerating over $F(\mathcal{S})\times\left|\mathcal{M}_{k-1}^{\mathrm{vio}}\right|$ choices of $(f,m)$. The total number of choices is bounded from above by $F(\mathcal{S})\times\left|\mathcal{M}_{0}^{\mathrm{vio}}\right|\times T$
in the worst case, where $\sum_{m\in\mathcal{M}}\underline{T}_{m}\le T \le \sum_{m\in\mathcal{M}}\overline{T}_{m}$.
Since the cooperative beamforming problem for each choice of $(f,m)$ can be solved in polynomial time, the overall computational complexity of Algorithm \ref{alg1} grows only polynomially with the number of BSs. In general, the proposed greedy algorithm is suboptimal. However, in the special case of large cache capacity, Algorithm \ref{alg1} terminates without the need to solve \eqref{eq:greedy} and the obtained solution is globally optimal. 

\begin{algorithm}[t]
\protect\caption{\textcolor{black}{Greedy Iterative Algorithm for Solving R0} }

\label{alg1} \begin{algorithmic}[1]
\STATE \textbf{Initialization}: $\mathcal{Q}_{0}\leftarrow\prod_{f\in\mathcal{F}(\mathcal{S})}\mathcal{M}_{f}^{\mathrm{F-Coop}}$
, $k \leftarrow 1$;

\STATE Solve Problem R0($\mathbf{w}_{\boldsymbol{\rho}}$) according to $\mathcal{Q}_{0}$;

\WHILE{$\mathcal{M}_{k-1}^{\mathrm{vio} } \neq \emptyset$ (cf. (\ref{eq:vioBS}))} 

\FOR{\textbf{each} $(f,m) \in \mathcal{F}(\mathcal{S})\times\mathcal{M}_{k}^{\mathrm{vio}}$}

\STATE Solve Problem R0($\mathbf{w}_{\boldsymbol{\rho}}$) according to  $ \mathcal{Q}_{k-1} \backslash \left\{ (f,m)\right\} $;

\ENDFOR

\STATE $\mathcal{Q}_{k} \leftarrow \mathcal{Q}_{k-1}\backslash\left\{ (f',m')\right\}$, where $(f',m')$ solves (\ref{eq:greedy});

\STATE $k \leftarrow k+1$, 
\ENDWHILE
\end{algorithmic} 
\end{algorithm}

\subsection{Solution of Problem Q0\label{sec4-3}}
Problem Q0 has a considerably ($\Omega$-times) larger problem size than R0. Solving Problem Q0 via enumeration methods seems impossible due to the overwhelming computational complexity. Besides, the greedy suboptimal method (cf. Algorithm~\ref{alg1}) is not directly applicable for solving Q0 since constraint C1 becomes bilinear over the joint optimization space $\left\{ \mathbf{C}_{\text{\mbox{I}}}\right\} $ of Q0. Yet, we handle both issues by applying the following relaxation method. 

In particular, the bilinear constraint C1 is transformed to 
\begin{equation}
\widetilde{\textrm{C1}}:\left\{ \begin{array}{c}
c_{f,m}+b_{f,l,m,\omega}\ge q_{f,l,m,\omega},\\
c_{f,m},b_{f,l,m,\omega}\in[0,1],\; q_{f,l,m,\omega}\in\left\{ 0,1\right\} .
\end{array}\right.
\end{equation}
If the average backhaul capacity is insufficient, $\widetilde{\textrm{C1}}$ and C3 together lead to $b_{f,l,m,\omega}=(1-c_{f,m})\times q_{f,l,m,\omega}$ since Q0 enjoys a similar monotonicity as R0, cf. Lemma \ref{lem:(Monotonicity-of-R0)}; otherwise, $\widetilde{\textrm{C1}}$ is inactive. Thus, $\widetilde{\textrm{C1}}$ and C1 are equivalent.

Moreover, let $\widehat{\textrm{C1}}$ be a relaxation of $\widetilde{\textrm{C}1}$ where the binary constraints are replaced by $q_{f,l,m,\omega}\in[0,1]$. By adopting $\widehat{\textrm{C1}}$ in Q0, we arrive at 
\begin{align}
\textrm{Q1:}\quad\min_{\mathbf{C}_{\text{\mbox{I}}}}\quad & \frac{1}{\Omega}\sum\nolimits_{\omega=1}^{\Omega}\; f_{\mathrm{\text{\mbox{I}}},\omega}\\
\mathrm{s.t.}\quad & \textrm{C2, }\ensuremath{\overline{\textrm{C3}}},\;\mathbf{D}_{\text{\mbox{I}},\omega}\in\widehat{\mathcal{D}}_{\text{\mbox{I}},\omega},\;\omega\in\left\{ 1,\ldots,\Omega\right\} ,\nonumber 
\end{align}
where $\widehat{\mathcal{D}}_{\text{\mbox{I}},\omega}\triangleq\left\{ \mathbf{D}_{\text{\mbox{I}},\omega}\mid\widehat{\textrm{C1}}\textrm{, C4--C7}\right\} $.
Problem Q1 remains non-convex. However, the hidden convexity of Problem Q1 can be shown in a similar manner as that of R0 in Theorem \ref{prop2}. Thus, the relaxed problem can be solved  efficiently (in polynomial time) via the interior point method \cite{Bertsekas82Lagrange}. Moreover, the following theorem establishes that the solution of the relaxed problem is asymptotically optimal for sufficiently large $\Omega$.

\begin{thm}
\label{thm2}
\emph{Problems Q1 and Q0 are equivalent as $\Omega \to \infty$ in the sense that  their optimum value and optimal caching decisions are the same.}
\end{thm}
\begin{IEEEproof}
Please refer to Appendix~\ref{append}.
\end{IEEEproof}

\section{\label{sec:Simulation-Results}Simulation Results}

\begin{table}[t]
\centering\protect\caption{\label{tab1}Simulation parameters.}
\small
\begin{tabular}{|c|c|}
\hline 
Parameters & Settings\tabularnewline
\hline 
\hline 
System bandwidth & 10 MHz\tabularnewline
\hline
Duration of time slot &$\tau =$ 10 ms\tabularnewline  
\hline 
File splitting & $L=45 \textrm{ min} / \tau =2.7 \times 10^4$ \tabularnewline  
\hline 
BS transmit power & $P_{m}^{\max}=$ 46 dBm\tabularnewline
\hline 
Noise power density & $-$172.6 dBm/Hz\tabularnewline
\hline 
Delivery QoS requirement & $R_{\boldsymbol{\rho}}^{\textrm{req}}=1.1 \times R_{\boldsymbol{\rho}}^{\mathrm{sec}} =$ 1.65 Mbps \tabularnewline
\hline 
Delivery secrecy threshold & $R_{\mathrm{e},\boldsymbol{\rho}}^{\mathrm{tol}}= 0.1\times R_{\boldsymbol{\rho}}^{\textrm{sec}}=$ 150 kbps\tabularnewline
\hline 
\end{tabular}
\end{table}

In this section, we evaluate the system performance for the proposed caching and secure delivery schemes. Consider a cluster of \textbf{$M=7$} hexagonal cells, where a BS is deployed at the center of each cell with an inter-BS distance of 500~m. Each BS is equipped with $N_{\mathrm{t}}=4$ antennas and the ER has $N_{\mathrm{e}}=2$ antennas. We assume that $F=$ 10 video files, each of duration 45 minutes and size 500 MB (Bytes), are delivered to $K=5$ single-antenna LRs. Consequently, an estimated secrecy data rate of $R_{\boldsymbol{\rho}}^{\mathrm{sec}} =  Q_{f} = 500 \times 8.0\times10^{6}/(45\times60)\approx1.5$~Mbps is required at each LR for secure and uninterrupted video streaming.  The LRs and the ER are uniformly and randomly distributed in the system while the minimum distance between receiver and BS is 50 m. Each LR requests one file independent of the other LRs. Let $\theta_{f}$ be the probability of file $f\in\mathcal{F}$ being requested and $\boldsymbol{\theta}=[\theta_{1},\ldots,\theta_{F}]$ be the probability distribution of the requests for the different files. We set $\theta_{f}=\frac{1}{f^{\kappa}}/\sum_{f\in\mathcal{F}}\frac{1}{f^{\kappa}}$ with $\kappa=1.1$ according to the Zipf distribution \cite{Breslau99Zipf}. Moreover, the 3GPP path loss model (``Urban Macro NLOS'' scenario) in \cite{3GPP:TR36814} is adopted. The capacities of the backhaul links
are independently and identically distributed (i.i.d.) as $\mathsf{Pr}(B_{m}^{\max}=0 \textrm{ Mbps})=0.3$, $\mathsf{Pr}(B_{m}^{\max}=3 \textrm{ Mbps})=0.4$, and $\mathsf{Pr}(B_{m}^{\max}=6 \textrm{ Mbps})=0.3$,
$\forall m$, which can be interpreted as the probabilities of high, medium, and low non-VoD traffic scenarios in the cellular network, respectively. The other relevant system parameters are given in Table \ref{tab1}. Before video delivery starts, $\Omega=50$ scenarios are randomly generated based on the models for user preference, CSI, and backhaul capacity  to train the initial cache status, cf. Problem Q0.

As a performance benchmark, the optimal solution of R0 is evaluated by exhaustive search. Besides, the following caching and delivery schemes are considered as baseline schemes:
\begin{itemize}
\item Baseline 1 (Preference-based caching): The most popular files are cached. Assuming $\boldsymbol{\theta}$ is known, the cache control decision is made based on
\begin{equation*}
\begin{aligned}\max_{c_{f,m}} & \quad\sum\nolimits _{f,m}{\theta_{f}  c_{f,m}  V_{f}}\\
\textrm{s.t}. & \quad c_{f,m}\in[0,1],\;\textrm{C2}.
\end{aligned}
\end{equation*}

\item Baseline 2 (Uniform caching): The same amount of data is cached for each file, i.e., 
\[c_{f,m} V_{f} =\frac{1}{F} \times \min\{C_{m}^{\max},\,\sum\nolimits _{f=1}^{F}V_{f}\}, \; \forall f,m, \]
and the user's preference is not taken into account. For Baselines 1 and 2, the proposed delivery scheme, i.e., Algorithm~\ref{alg1}, is adopted. 

\item Baseline 3 (Coordinated beamforming): The user is associated with one of the nearest BSs which has sufficient backhaul capacity available. Each video (sub)file is only delivered from the associated BS, i.e., $\sum_{m\in\mathcal{M}}q_{f,l,m}=1,\forall (f,l)\in\mathcal{F}\times\mathcal{L}$. 

\item Baseline 4 (Full BS cooperation): The backhaul capacity constraints are dropped and all BSs cooperate to serve all users, i.e., $q_{f,l,m}=1,\forall f,l,m$. For Baselines 3 and 4, the optimal beamforming solutions are obtained based on R0($\mathbf{D}_{\text{\mbox{II}},2}$) in which the $\{q_{f,l,m}\}$ are fixed accordingly. 
\end{itemize}

\begin{figure}[t]
\vspace{-.4cm}
\centering 
\subfloat[]{\includegraphics[scale=0.35]{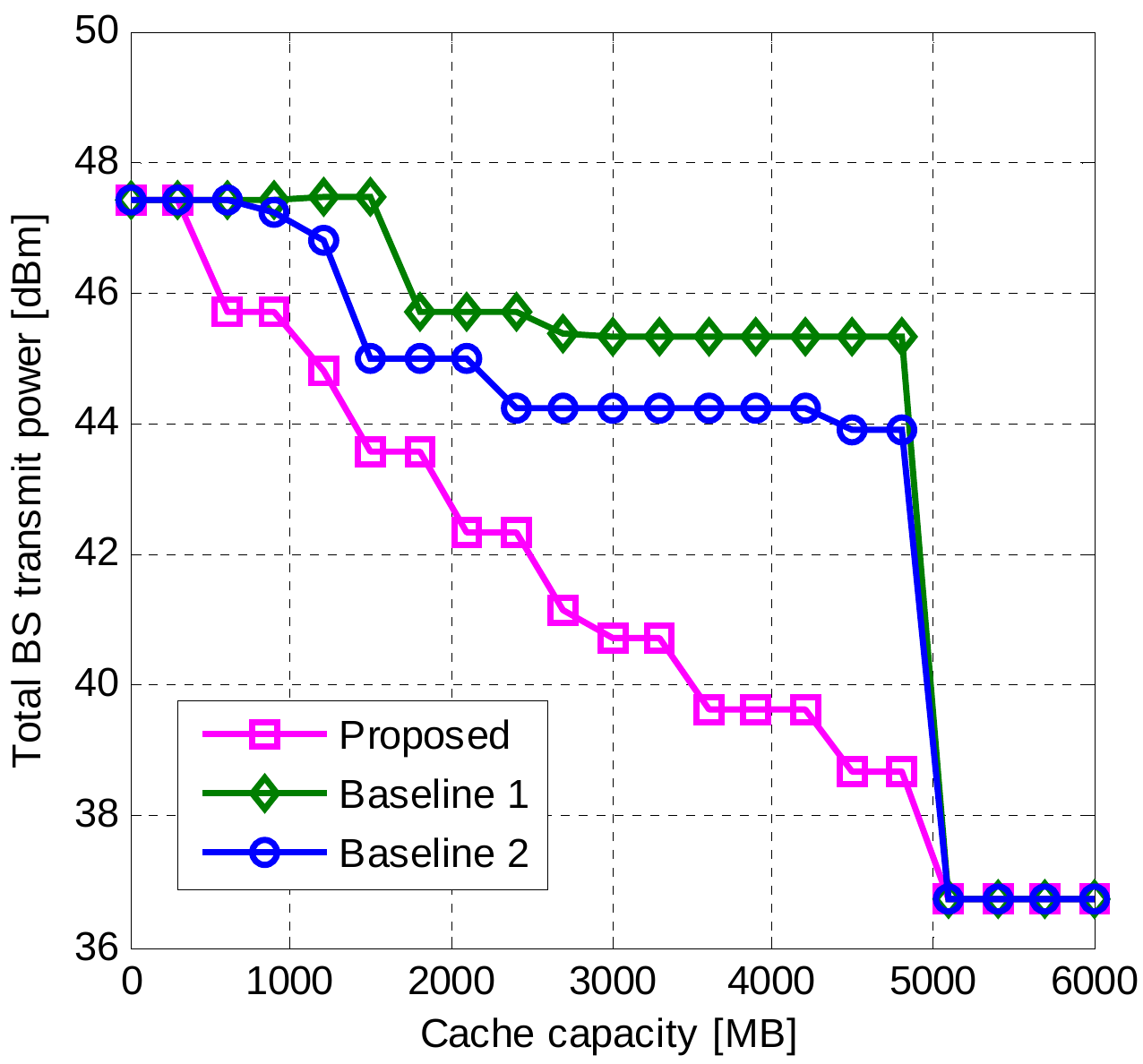} \label{fig:cach}} 
\subfloat[]{\includegraphics[scale=0.35]{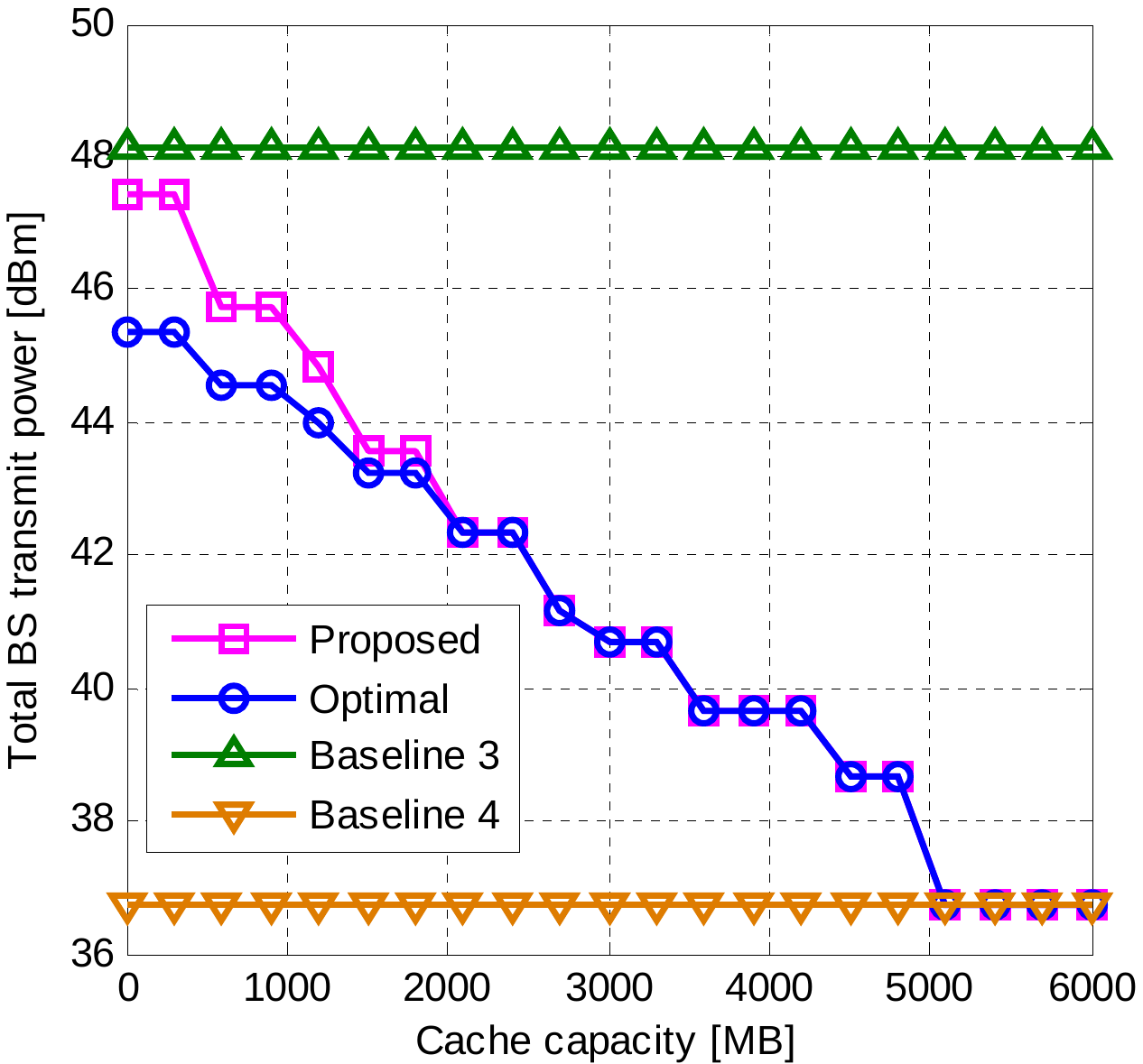} \label{fig:del}}
\protect\caption{Total BS transmit power versus cache capacity. }
\vspace{-.25cm}
\end{figure}

Figures~\ref{fig:cach} and \ref{fig:del} illustrate the performances of the considered caching and delivery schemes as functions of the cache capacity, respectively. The initial cache status in Figure~\ref{fig:del} is determined based on Q0. As can be observed from Figure~\ref{fig:cach}, a larger cache capacity leads to a lower total BS transmit power as larger (virtual) transmit antenna arrays can be formed during video delivery. For example, the average number of cooperating BSs for the proposed scheme is $4.0$ for $C_m^{\max} = 4000$ MB compared to $2.6$ for $C_m^{\max} = 1000$ MB, which leads to a transmit power reduction of up to $6$ dB. The performance gap between the considered caching schemes is only negligible for small (large) cache capacities because of insufficient (saturated) BS cooperation. For medium cache capacities, however, the proposed caching scheme achieves considerable transmit power savings due to its ability to exploit historical information of user requests, the backhaul, and the CSI. Note also that, when the increase in cache capacity is insufficient to support additional BSs for cooperative transmission, the performance remains constant. As a result, the total transmit power decreases in a piece-wise constant manner with increasing cache capacities.

From Figure~\ref{fig:del} we observe that, as expected, Baselines 3 and 4 constitute performance  lower and upper bounds for the proposed delivery scheme, respectively. Comparing the proposed delivery scheme and the optimal delivery scheme (involving an exhaustive search), the performance gap between them reduces as the cache capacity increases. This is because, for large cache capacities, less backhaul traffic is created and correspondingly the possibility of C3 being active is reduced. It is interesting to observe that for a cache capacity of  2000~MB, the proposed scheme already achieves the optimal performance.

\begin{figure}[t]
\vspace{-.4cm}
\centering
\subfloat[]{\includegraphics[scale=0.35]{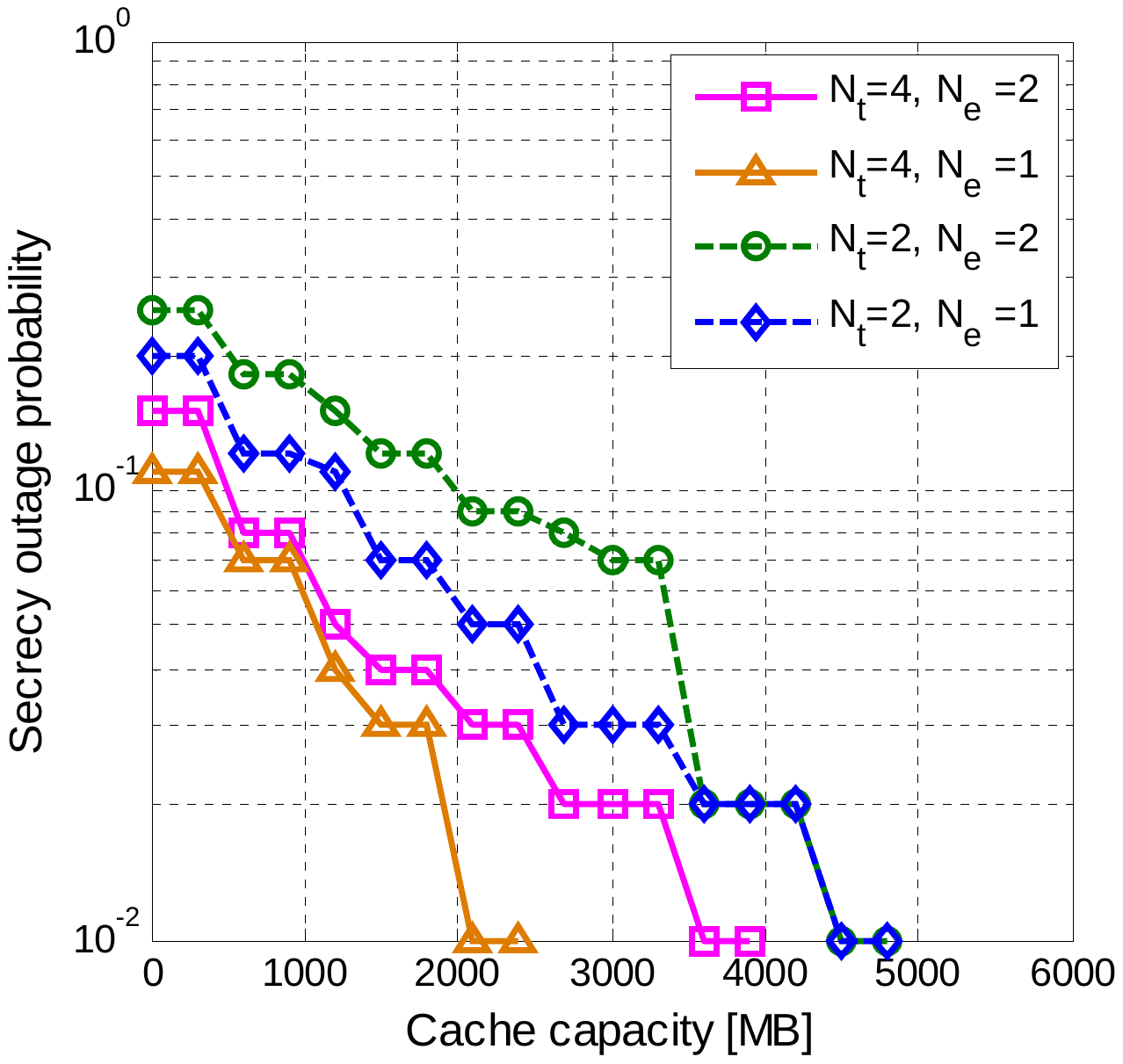} \label{fig:ant1}}  
\subfloat[]{\includegraphics[scale=0.35]{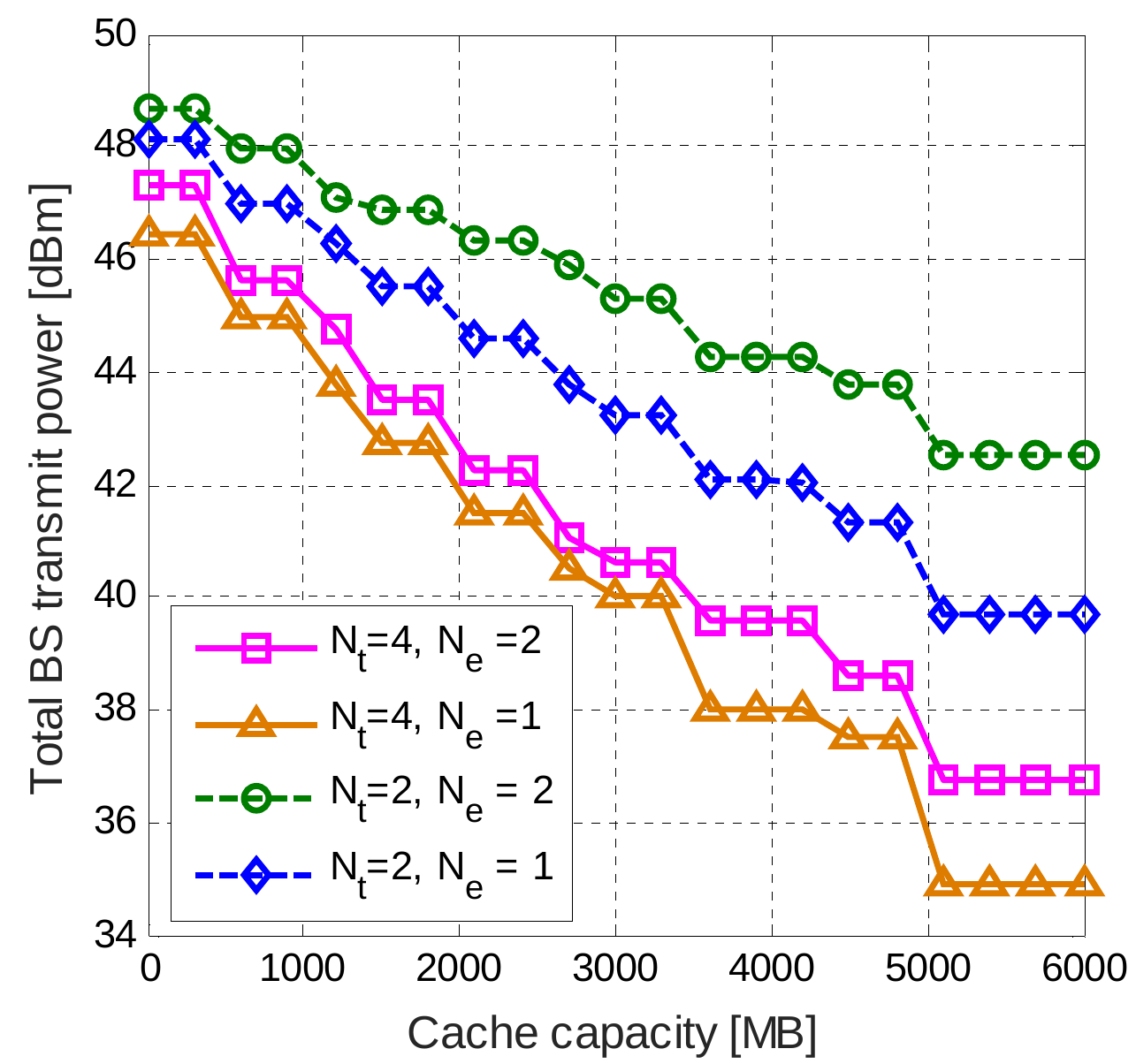} \label{fig:ant2}}
\protect\caption{(a) Secrecy outage probability and (b) total BS transmit power of proposed delivery scheme versus cache capacity for different numbers of antennas.}
\end{figure}

In Figures~\ref{fig:ant1} and \ref{fig:ant2}, the secrecy outage probability, defined as $p_{\textrm{out}} \triangleq \mathsf{Pr}(R_{\boldsymbol{\rho}}^{\mathrm{sec}} < [R_{\boldsymbol{\rho}}^{\textrm{req}}-R_{\mathrm{e},\boldsymbol{\rho}}^{\mathrm{tol}}]^{+})$, and the total BS transmit power of the proposed delivery scheme are evaluated for different numbers of transmit and eavesdropper antennas, respectively. Herein, $p_{\textrm{out}}$ characterizes the likelihood that Problem R0 is infeasible because either the QoS constraint C6 or the secrecy constraint C7 fails to be satisfied. For a given cache capacity, larger $N_{\rm{t}}$ at the BSs and smaller $N_{\rm{e}}$, both directly contributing to increased s.d.o.f., lead to a lower secrecy outage probability as well as transmit power savings, cf. Figures~\ref{fig:ant1} and \ref{fig:ant2}, respectively. A further reduction in the secrecy outage probability and the total BS transmit power can be achieved by increasing the cache capacity at the BSs. 

\vspace{-.2cm}
\section{\label{sec:Conclusion}Conclusion}
In this paper, caching was exploited as a physical layer security mechanism for cellular video streaming, where caching can reduce the backhaul capacity required for cooperative transmission among large groups of BSs and thereby increase the available secure degrees of freedom. Caching and cooperative transmission were optimized based on a mixed-integer two-stage problem. As the problem is NP-hard, suboptimal polynomial time algorithms were developed to solve the problem efficiently. The optimality of the proposed algorithms was verified in the regimes of large cache capacity and large numbers of training scenarios, respectively. Simulation results showed that the proposed caching and delivery schemes can significantly enhance both the physical-layer security and power efficiency of cellular video streaming. This paper assumed perfect knowledge of the ER's CSI to obtain a performance upper bound. The case of imperfect CSI will be considered in future work.

\vspace{-.2cm}
\appendices{}
\section{\label{append2}Proof of Theorem \ref{prop2}}
\vspace{-.1cm}
Note that R0($\mathbf{w}_{\boldsymbol{\rho}}$) and R1 are equivalent
if and only if the rank constraint $\mathsf{rank}(\mathbf{W}_{\boldsymbol{\rho}}^{*})\le1$
is fulfilled. To this end, let $\boldsymbol{\alpha}=[\alpha_{m\boldsymbol{\rho}}]$,
$\boldsymbol{\beta}=[\beta_{m}]$, $\boldsymbol{\lambda}=[\lambda_{\boldsymbol{\rho}}]$,
$\boldsymbol{\Phi}_{\boldsymbol{\rho}}$, and $\boldsymbol{\Theta}_{\boldsymbol{\rho}}=[\boldsymbol{\Theta}_{1\boldsymbol{\rho}},\,\boldsymbol{\Theta}_{2\boldsymbol{\rho}}]$
be the Lagrangian multipliers associated with constraints $\overline{\textrm{C4}}$,
$\overline{\textrm{C5}}$, $\overline{\textrm{C6}}$, $\overline{\textrm{C7}}$,
and $\overline{\textrm{C8}}$, respectively, where
\begin{alignat*}{1}
&\alpha_{m\boldsymbol{\rho}}\ge0,\beta_{m}\ge0,\lambda_{\boldsymbol{\rho}}\ge0,\boldsymbol{\Phi}_{\boldsymbol{\rho}}\succeq\mathbf{0},
\boldsymbol{\Theta}_{1\boldsymbol{\rho}}\succeq\mathbf{0},\textrm{and }\boldsymbol{\Theta}_{2\boldsymbol{\rho}}\succeq\mathbf{0}.
\end{alignat*}
 The Lagrangian of Problem R1 is formulated as
\begin{equation*}
\begin{aligned} & \mathcal{L}(\mathbf{W}_{\boldsymbol{\rho}};\,\boldsymbol{\alpha},\boldsymbol{\beta},\boldsymbol{\lambda},\boldsymbol{\Phi}_{\boldsymbol{\rho}},\boldsymbol{\Theta}_{\boldsymbol{\rho}})\\
 & = \mathsf{tr}\left[\sum\nolimits_{\boldsymbol{\rho}}\left(\mathbf{B}_{\boldsymbol{\rho}}
-2\lambda_{\boldsymbol{\rho}}\mathbf{H}_{\boldsymbol{\rho}}
-\boldsymbol{\Theta}_{1\boldsymbol{\rho}}\right)\mathbf{W}_{\boldsymbol{\rho}}\right]
+\Delta,
\end{aligned}
\label{eq:Lag_Thm2}
\end{equation*}
where $\Delta$ is a collection of terms irrelevant for the proof and 
\begin{equation*}
\begin{aligned}\mathbf{B}_{\boldsymbol{\rho}} & \triangleq\mathbf{I}+\boldsymbol{\Lambda}_{\boldsymbol{\rho}}^{\boldsymbol{\alpha},\boldsymbol{\beta}}+\mathbf{G}\boldsymbol{\Phi}_{\boldsymbol{\rho}}\mathbf{G}^{H}+\sum\nolimits_{\boldsymbol{\rho}\in\mathcal{S}}(1+\kappa_{\boldsymbol{\rho}}^{\mathrm{req}})\lambda_{\boldsymbol{\rho}}\mathbf{H}_{\boldsymbol{\rho}}\succ\mathbf{0},\end{aligned}
\label{eq:Lag_Thm2_B}
\end{equation*}
with $\boldsymbol{\Lambda}_{\boldsymbol{\rho}}^{\boldsymbol{\alpha},\boldsymbol{\beta}} \triangleq \sum_{m \in \mathcal{M}}  (\alpha_{m\boldsymbol{\rho}}+\beta_{m} ) \boldsymbol{\Lambda}_{m}$ and 
$\boldsymbol{\Lambda}^{\boldsymbol{\beta}}\triangleq\sum_{m\in\mathcal{M}}\beta_{m}\boldsymbol{\Lambda}_{m}$.
Note that R1 is a convex optimization problem fulfilling Slater's
condition. Thus, strong duality holds for R1 and the Karush\textendash Kuhn\textendash Tucker
(KKT) conditions are both necessary and sufficient for a primal-dual
point $(\mathbf{W}_{\boldsymbol{\rho}};\,\boldsymbol{\alpha},\boldsymbol{\beta},\boldsymbol{\lambda},\boldsymbol{\Phi}_{\boldsymbol{\rho}},\boldsymbol{\Theta}_{\boldsymbol{\rho}})$
to be optimal. The KKT conditions for R1 are given by 
\begin{alignat}{1}
\nabla_{\mathbf{W}_{\boldsymbol{\rho}}}\mathcal{L}=\mathbf{B}_{\boldsymbol{\rho}}-2\lambda_{\boldsymbol{\rho}}\mathbf{H}_{\boldsymbol{\rho}}-\boldsymbol{\Theta}_{1\boldsymbol{\rho}}=\mathbf{0},\label{eq:kkt1}\\
\mathbf{W}_{\boldsymbol{\rho}}\boldsymbol{\Theta}_{1\boldsymbol{\rho}}=\mathbf{0},\label{eq:kkt2}\\
\mathbf{W}_{\boldsymbol{\rho}}\succeq\mathbf{0},\quad\lambda_{\boldsymbol{\rho}}\ge0.\label{eq:kkt3}
\end{alignat}
Based on \eqref{eq:kkt1} and \eqref{eq:kkt2}, we have $\mathbf{W}_{\boldsymbol{\rho}}\mathbf{B}_{\boldsymbol{\rho}}=2\lambda_{\boldsymbol{\rho}}\mathbf{W}_{\boldsymbol{\rho}}\mathbf{H}_{\boldsymbol{\rho}}$.
Since $\mathsf{rank}(\mathbf{H}_{\boldsymbol{\rho}}) \le1$, the rank of the optimal $\mathbf{W}_{\boldsymbol{\rho}}$ can be determined as 
\begin{alignat*}{1}
\mathsf{rank}(\mathbf{W}_{\boldsymbol{\rho}}) & \stackrel{\textrm{(a)}}{=}\mathsf{rank}(\mathbf{W}_{\boldsymbol{\rho}}\mathbf{B}_{\boldsymbol{\rho}})\stackrel{\textrm{(b)}}{=}\mathsf{rank}(\lambda_{\boldsymbol{\rho}}\mathbf{W}_{\boldsymbol{\rho}}\mathbf{H}_{\boldsymbol{\rho}})\\
 & \stackrel{\textrm{(c)}}{\le}\min\left\{ \mathsf{rank}(\lambda_{\boldsymbol{\rho}}\mathbf{W}_{\boldsymbol{\rho}}),\,\mathsf{rank}(\mathbf{H}_{\boldsymbol{\rho}})\right\} \le1,
\end{alignat*}
where (a) is due to $\mathbf{B}_{\boldsymbol{\rho}}\succ\mathbf{0}$,
(b) is a result of \eqref{eq:kkt1} and \eqref{eq:kkt2}, and (c)
follows from the rank inequality $\mathsf{rank}(\mathbf{AB})\le\min\left\{ \mathsf{rank}(\mathbf{A}),\,\mathsf{rank}(\mathbf{B})\right\} $.
Thus, $\mathsf{rank}(\mathbf{W}_{\boldsymbol{\rho}})\le1$.
This completes the proof. 

\vspace{-.2cm}
\section{\label{append}Proof of Theorem~\ref{thm2}}

We first assume that the caching decisions in Problems Q0 and Q1 are given, where the resulting problems are denoted by Q0($\mathbf{D}_{\text{\mbox{I}},\omega}$) and Q1($\mathbf{D}_{\text{\mbox{I}},\omega}$), respectively. Note that owing to $\ensuremath{\overline{\textrm{C3}}}$, Q0($\mathbf{D}_{\text{\mbox{I}},\omega}$) falls into a class of separable integer programming problems \cite[Chapter 5.6]{Bertsekas82Lagrange}. Moreover, based on \cite[Proposition 5.26]{Bertsekas82Lagrange}, for such problems, there is only a negligible difference between the primal and the dual optimal values for large values of $\Omega$, i.e., 
\begin{equation}
\lim_{\Omega\to\infty}(f^{*}-q^{*})=0,
\label{eq:nullgap}
\end{equation}
where $f^{*}$ and $q^{*}$ are the primal and the dual optimal values of Q0($\mathbf{D}_{\text{\mbox{I}}, \omega}$), respectively. Please refer to \cite[Proposition 5.26]{Bertsekas82Lagrange} for a proof of \eqref{eq:nullgap}. 

Meanwhile,  Q1($\mathbf{D}_{\text{\mbox{I}},\omega}$) is a relaxed version of Q0($\mathbf{D}_{\text{\mbox{I}},\omega}$), where the binary constraints are relaxed into affine ones. Due to the convexity of Q1, strong duality holds for Problem Q1($\mathbf{D}_{\text{\mbox{I}},\omega}$). Based on Lagrangian duality theory, it can be further shown  that the dual problems of Q1($\mathbf{D}_{\text{\mbox{I}},\omega}$) and Q0($\mathbf{D}_{\text{\mbox{I}},\omega}$) are identical. Consequently, the optimal value of Q1($\mathbf{D}_{\text{\mbox{I}},\omega}$) is also given by $q^*$. According to \eqref{eq:nullgap}, the difference between the optimal values of Q1($\mathbf{D}_{\text{\mbox{I}},\omega}$) and Q0($\mathbf{D}_{\text{\mbox{I}},\omega}$) then becomes negligible for sufficiently large $\Omega$. Since \eqref{eq:nullgap} holds for arbitrary caching decisions, the performance gap between  Q1 and Q0 also vanishes as $\Omega\to\infty$, which completes the proof. 

\vspace{-.1cm}
\bibliographystyle{IEEEtran}
\bibliography{IEEEabrv,bib/SecureCaching,bib/MIMO-IBC,bib/PLS,bib/OptimizationRefs,bib/WirelessCaching,bib/CoMP,bib/WirelessVideo}

\end{document}